\documentclass[conference,twocolumn]{IEEEtran}
\usepackage{fixltx2e}
\usepackage[latin1]{inputenc}
\usepackage{amsmath}
\usepackage{amsfonts}
\usepackage{amssymb}
\usepackage{slashbox}
\usepackage{pict2e}
\usepackage{epsfig}
\usepackage{psfrag}
\usepackage{array}
\usepackage{verbatim}
\usepackage{threeparttable}
\usepackage{algorithm}
\usepackage{algorithmic}
\newtheorem{theorem}{Theorem}

\newtheorem{prop}{Proporsition}
\newtheorem{problem}{Problem}
\renewcommand{\mag}{\text{Mag}}
\newcommand{\thetabm}{ {\boldsymbol{\theta}} }
\newcommand{\deltabm}{ {\boldsymbol{\delta}} }
\newcommand{\omegabm}{ {\boldsymbol{\omega}} }

\newcommand{\Utheta}{ U_{\thetabm}}

\newcommand{\supp}{ \text{supp} }
\newcommand{\emphb}{\textbf}
\newcommand{\RealD}{\mathbb{R}}
\newcommand{\stepzero}{\emphb{Step 0}}
\newcommand{\stepone}{\emphb{Step 1}}
\newcommand{\steptwo}{\emphb{Step 2}}
\newcommand{\stepthree}{\emphb{Step 3}}
\newcommand{\stepfour}{\emphb{Step 4}}
\newcommand{\stepfive}{\emphb{Step 5}}

\newcommand{\bec}{}

\begin{document}
\IEEEoverridecommandlockouts
\title{A Bio-Inspired Robust Adaptive Random Search Algorithm for Distributed Beamforming}
\author{\IEEEauthorblockN{Chia-Shiang Tseng, ~Chang-Ching Chen, ~Che Lin}
\thanks{This work is supported by National Science Council, Taiwan (R.O.C.), under Grant NSC99-2221-E-007-089-MY3.}
\IEEEauthorblockA{Institute of Communication Engineering \& Department of Electrical Engineering\\
National Tsing Hua University, Hsinchu, Taiwan 30013 \\
E-mail: s9964518,~s9864511,~clin@ee.nthu.edu.tw}
}

\maketitle
\noindent \begin{abstract}
A bio-inspired robust adaptive random search algorithm (BioRARSA), designed for distributed beamforming for sensor and relay networks, is proposed in this work.
It has been shown via a systematic framework that BioRARSA converges in probability and its convergence time scales linearly with the number of distributed transmitters. More importantly, extensive simulation results demonstrate that the proposed BioRARSA outperforms existing adaptive distributed beamforming schemes by as large as 29.8\% on average. This increase in performance results from the fact that BioRARSA can adaptively adjust its sampling stepsize via the ``swim'' behavior inspired by the bacterial foraging mechanism. Hence, the convergence time of BioRARSA is insensitive to the initial sampling stepsize of the algorithm, which makes it robust against the dynamic nature of distributed wireless networks.
\end{abstract}

\section{Introduction}\label{sec:intro}
In recent years, the advancement of microelectronic technology makes possible the manufacture of low-cost transceivers that can be deployed with ease to form distributed sensor and relay networks. Since then, studies in distributed network have attracted enormous attentions.
However, there are innate limitations, e.g., the lack of constant power supply for distributed transmitters, that need to be particularly addressed in the applications of such networks.
A viable solution to overcome this energy scarcity is for the distributed transmitters to form an antenna array and transmit collaboratively using distributed beamforming. The problem remains, however, as the beamforming coefficients for each distributed transmitters cannot be easily obtained without complete knowledge of the channel state information (CSI), which is in general not available at the transmitter end.

A simple yet efficient adaptive distributed beamforming scheme that requires only a single bit feedback was proposed by Mudumbai et al in \cite{Mudumbai2006}. The convergence analysis has been conducted using different approaches in the literature \cite{Mudumbai2006,Lin2006,Bucklew2008,Chen2010}, where the  fast convergence and the linear scalability of the scheme have been well established.
With these extensive analyses, it has been shown that the simple one-bit adaptive scheme provides exceptional performance. The challenge ahead is then to discover new schemes that can further improve the performance without sacrificing much on the complexity.

In \cite{Lin2006}, we reformulated this adaptive distributed beamforming scheme as a local random search algorithm under a systematic framework.
Once reformulated in this random search framework, an immediately observation is that the performance can be improved by introducing \emph{Reverse Tracking}\footnote{Reverse tracking is based on the intuition that if moving in the forward direction decreases the objective function, it is highly possible that moving in the reverse direction improves the objective function.},
a rather old technique in the field of random search algorithms \cite{Schrack1976,Solis1981}.
Rigorous analysis has been carried out in \cite{Bucklew2008} to show that the convergence rate can indeed be doubled once this intuitive idea is applied to the one-bit adaptive scheme.
However, the exceptional performances of both the one-bit adaptive scheme and the improved reverse tracking scheme highly rely on proper tuning of system parameters with respect to the total number of transmitters, which may not be known a prior due to environmental variations or random node failures.

Surprisingly, the key to achieve breakthrough sometimes lies in a completely different field. In \cite{Passino2002}, a heuristic random search mechanism, that mimics the foraging behavior of E. Coli bacteria, was proposed. The randomized search strategy seems suitable for the phase alignment process of adaptive distributed beamforming, yet some features cannot be included due to the innate limitations of distributed networks. Therefore, we carefully redesign the bacterial foraging algorithm and
propose a bio-inspired robust adaptive random search algorithm (BioRARSA) that can adaptively adjust its sampling stepsize through the ``swim'' behavior of E. coli bacteria. As a result, BioRARSA is robust against all initial parameters and adapts well to the dynamic nature of distributed networks.

We organize the paper as follows:
In section~\ref{sec:syssetup}, the problem of adaptive distributed beamforming in distributed wireless networks is described,
and its connection to a random search framework is established.
The description of our proposed scheme, BioRARSA, is presented in section~\ref{sec:algdesc}. Furthermore, the convergence analysis using the analytical tools proposed in \cite{Chen2010} is provided in section~\ref{sec:converge}.
In section~\ref{sec:sim}, we show numerical simulations to demonstrate some interesting characteristics of BioRARSA. Finally, we conclude the advantages of BioRARSA and touch upon future works in section~\ref{sec:con}.

\section{Problem Formulation}\label{sec:syssetup}
In this section, we introduce the problem of adaptive distributed beamforming. To systematically analyze such problem, we further describe a recently proposed random search framework \cite{Chen2010} that greatly facilitate the convergence analysis.

\subsection{Distributed Beamforming with Limited Feedback}
We begin with the formulation of the distributed beamforming problem.
In our setting, $N_{s}$ distributed transmitters seek to transmit a common message $s \in \mathbb{C}$, with power constraint $\mathrm{E} \left [ |s|^{2} \right ]\leq P$, to the remote receiver.
Each transmitter and the receiver is equipped with a single antenna, where the channel between each transmitter and the receiver is assumed to be frequency flat and slow fading.
Also, each transmitter is allowed to alter its beamforming coefficient to achieve phase alignment.
We assume that the distributed transmitters can only cooperate via a one-bit, error-free reverse feedback link from the receiver, since local cooperation requires sophisticated media access control (MAC) layer protocols,  which may result in large energy dissipation and is not desirable for low-complexity distributed networks.

Based on the system specification, the discrete-time, complex baseband model over one coherence interval is given by
\begin{equation}\label{eq:model}
y[n] = \sum_{i=1}^{N_s} h_i g_i[n] s + w[n] = \sum_{i=1}^{N_s} a_i b_i[n] e^{j(\phi_i + \psi_i[n])} s + w[n]
\end{equation}
where $y[n] \in \mathbb{C}$ is the received signal, $h_{i} = a_i e^{j \phi_i}\in \mathbb{C}$ is the channel fading gain, $g_{i}[n] = b_{i}[n]e^{\psi_{i}[n]} \in \mathbb{C}$. is the beamforming coefficients, and $w[n] \sim {\cal CN}(0,\sigma^2)$ corresponds to the additive white Gaussian noise.

To simplify notations\footnote{The assumptions aim only for notation simplicity. As has been reported in \cite{Chen2010}, random search framework can be used to handle systems with more complicated power constraints, e.g., individual/total power constraints.}, we let $s = \sqrt{P}$ and impose fixed power constraint $b_{i} = 1$ among transmitters.
Assume that the magnitude of the signal component can be perfectly estimated, the received signal magnitude can be expressed as
\begin{equation}\label{eq:mag}
\mag(\theta_1[n],\cdots,\theta_{N_s}[n]) = \sqrt{P} \left | \sum_{i=1}^{N_s} a_i e^{j\theta_i[n]} \right |
\end{equation}
where $\theta_i[n] = \phi_i+\psi_i[n]$ is the total phase of the received signal from transmitter~$i$ at the $n$-th transmission.

In order to minimize detection error, it is clear we should maximize this magnitude function, and its global maxima are achieved when all phases are aligned, i.e., $\theta_{1}[n] = \theta_{2}[n] = \cdots = \theta_{N_{s}}[n]$.
In this setting, distributed transmitters seek to achieve this phase alignment adaptively with the help of a low rate feedback link. Consequently, we term such problem as adaptive distributed beamforming.

\subsection{A Random Search Framework}\label{sec:randomsearch}
With no prior knowledge of the channel fading coefficients, the received signal magnitude function is unknown for our system.
However, for the $n$-th transmission, the remote receiver obtains a sample of $\mag(\cdot)$ at $\thetabm[n] = \left[\theta_1[n] ~\cdots~ \theta_{N_s}[n] \right]$.
As a result, we can model the adaptive distributed beamforming problem as follows.
\begin{problem}
Given an unknown function $f : \Theta \rightarrow \RealD$ where only samples of $f(\cdot)$ is available, determine the global maximum value of $f(\cdot)$.\label{prob:rsa}
\end{problem}

For such problem, randomized schemes that utilize only samples of $\mag(\cdot)$ seems to be viable solutions.
To analyze the convergence behavior of such randomized schemes for Problem~\ref{prob:rsa}, we introduce the general random search framework proposed in \cite{Chen2010}. The framework describes a general random search algorithm, which we restate as follows:
\begin{itemize}
\item \stepzero: Initialize the algorithm by choosing an arbitrary initial point $\thetabm[0] \in \Theta.$
\item \stepone: Generate a random perturbation $\deltabm[n]$ according to a probability measure $\mu_n$, that could be time-varying.
\item \steptwo: Update $\thetabm[n]$ by $\thetabm[n] = D(\thetabm[n-1],\deltabm[n])$, where the mapping $D(\cdot,\cdot)$ satisfies the condition that $f(D(\thetabm[n-1],\deltabm)) \geq f(\thetabm[n-1])$.
\end{itemize}

Clearly, the study of random search algorithms is a much more mature field in the mathematical society \cite{Schrack1976,Solis1981}.
If the randomized scheme for adaptive distributed beamforming can be reformulated as the above random search algorithm, then the framework becomes a powerful tool that allows us to analyze the adaptive distributed beamforming schemes systematically.

The analysis of the one-bit scheme proposed by Mudumbai et al is an immediate example of the application of such framework as has been shown in \cite{Lin2006}, where
$\Theta$ is equal to $\RealD^{N_{s}}$, and $f(\cdot)$ is the function $\mag(\cdot)$. The probability measure in this case is uniformly distributed in a hypercube centered at the origin with sides equal to $2\Delta_{0}$. The function $D(\cdot,\cdot)$ is described by
\begin{align}
\nonumber D(\thetabm[n-1],\deltabm[n]) &=\\
 \thetabm[n-1] + &\deltabm[n] \cdot 1_{\{ \mag(\thetabm[n-1] + \deltabm[n]) > \mag(\thetabm[n-1])\}}
\end{align}
where $1_{\{\cdot\}}$ is the indicator function. As a result, the one-bit scheme can be reformulated as a simple local random search with fixed sampling stepsize.

As has been shown in \cite{Mudumbai2006,Lin2006,Chen2010}, the simple local random search algorithm performs surprisingly well.
In this work, we hope to propose a new algorithm with even better performance, i.e, one that achieve phase alignment with less time or power,
based on further understanding of the problem via the  random search framework, and inspiration from the foraging behavior of E. coli bacteria in biology.

\section{Bio-inspired Robust Adaptive \\ Random Search Algorithm - BioRARSA}\label{sec:algdesc}
As indicated by its name, BioRARSA is inspired by the foraging behavior of E. coli bacteria.
In \cite{Passino2002}, the foraging strategy and the life cycle of E. coli bacteria have been studied extensively.
When foraging, a E. coli bacterium randomly chooses a direction and swims towards the chosen direction until the density of its food fails to improve.
This randomized search strategy has been shown to be effective and applied in many fields. This motivates the inclusion of the ``swim'' behavior into our algorithm. By careful incorporation of the foraging behavior into our algorithm, we propose BioRARSA that achieves superior performance and is highly robust against system uncertainty.

\subsection{Algorithm Description}

\begin{table} [t]
\caption{Algorithm I - BioRARSA}\label{tab:alg}
\label{alg:biorarsa}
\begin{tabular*}{\columnwidth}{c}
\hline
\hline
\end{tabular*}
\begin{algorithmic}
\STATE Initialize parameters
\REPEAT
\FOR{$i \leftarrow 1$ to $L.Helds$}
\STATE $\deltabm[k] \sim \text{uni}\left( [ -\Delta_k, \Delta_k ] ^{N_{s}}\right )$
\IF {$\mag(\thetabm[k-1] + \deltabm[k]) < \mag(\thetabm[k-1])$}
\STATE $\deltabm[k] \leftarrow -\deltabm[k] $
\ENDIF
\IF {$\mag(\thetabm[k] + \deltabm) > \mag(\thetabm[k-1])$}
\STATE $\thetabm[k] \leftarrow \thetabm[k] + \deltabm$
\STATE $ \beta_{k} \leftarrow 1$
\WHILE {$\mag(\thetabm[k] + \deltabm[k]) > \mag(\thetabm[k])$
\\ \quad \quad \emphb{and} $\beta_{k} < L.Swim$}
\STATE $\thetabm[k] \leftarrow \thetabm[k] + \deltabm[k]$
\STATE $ \beta_{k} \leftarrow \beta_{k} + 1$
\ENDWHILE
\ENDIF
\STATE $k \leftarrow k + 1$
\ENDFOR
\STATE $Avg.Swim \leftarrow \max\left ( \alpha , \sum_{i = k - L.Held + 1}^{k}\beta_{k} \ / \ {L.Helds} \right )$ 
\STATE $\Delta_{k+1} \leftarrow \Delta_k \cdot Avg.Swim$
\UNTIL{Stopping criteria reached}
\end{algorithmic}
\begin{tabular*}{\columnwidth}{c}
\hline
\hline
\end{tabular*}
\end{table}

Now we introduce the proposed algorithm in details. The pseudocode of proposed algorithm is presented as  Algorithm~\ref{alg:biorarsa} in Table~\ref{tab:alg}, and we elaborate the algorithm step-by-step as follows.

\emphb{Step 1 - Initialization}: We first initializes the system by generating the initial phase $\thetabm[0] \in \mathbb{R}^{N_{s}}$. Other parameters of the system, including $L.Held$, $L.Swim$, and $\Delta_{0}$, can be transmitted from the receiver to the transmitters or pre-coded into the system in this stage. The purpose of these parameters will become clear in the illustration of later steps.

\emphb{Step 2 - Random Perturbation}: Assume now that the system is at the $k$-th iteration. Each distributed transmitter in the system selects a random perturbation $\delta_{i}[k]$ that is uniformly distributed within $[-\Delta_k,  \Delta_k]$,
where $\Delta_k$ is the sampling stepsize of the algorithm and can be adjusted iteratively.
From the perspective of the entire system, a random vector $\deltabm[k] = \left [ \delta_{1}[k] \cdots \delta_{N_{s}}[k] \right ]^{T} $ that is uniformly distributed in $\left [-\Delta_{k}, \Delta_{k}\right ]^{N_{s}}$ is jointly generated by distributed transmitters.

Once the random perturbations are generated, distributed transmitters add the perturbations to their beamforming coefficients and transmit the common message $s$ simultaneously.
Upon receiving the composite signal $y[k]$, as in \eqref{eq:model}, the receiver compares the received signal magnitude with the largest magnitude recorded up to the $(k-1)$-th iteration, and broadcasts a one-bit feedback signal to the transmitters.
The broadcasted binary signal is given by
\begin{equation}\label{eq:rsscomp}
S = 1_{\left \{ \mag(\thetabm') \  > \ \underset{i = 1 , \cdots, k-1}{\max}\mag(\thetabm[i])\right \}}
\end{equation}
where $\thetabm' = \thetabm[k-1] + \deltabm[k]$.
If ``1'' is broadcasted, i.e., received signal magnitude increased with perturbations, the algorithm enters \stepfour. Otherwise, the algorithm enters \stepthree.

\emphb{Step 3 - Reverse tracking}:
Each distributed transmitter multiplies its random perturbation $\delta_{i}[k]$ generated in \steptwo \ by $-1$, and use it to perturb the beamforming coefficients for the next transmission.
After comparing the received signal magnitude with the previous ones, the receiver broadcasts a one-bit feedback signal back to transmitters as in \steptwo. The broadcasted signal is the same as in \eqref{eq:rsscomp}, except that now $\thetabm' = \thetabm[k-1] - \deltabm[k]$.
If ``1'' is broadcasted, the algorithm enters \stepfour. Otherwise, the algorithm enters \stepfive.

\emphb{Step 4 - Swim}:
As the perturbations generated in \steptwo, or their reversed versions in \stepthree \ increase the received signal magnitude, the algorithm will proceed to search along the same direction by adding the same perturbations to the beamforming coefficients.
Again, the received signal magnitude achieved by these perturbed beamforming coefficients is compared with the previous ones as in \steptwo \ and \stepthree.
If ``1'' is broadcasted, transmitters update the beamforming coefficients with the perturbed value and return to the beginning of \stepfour. Otherwise, the algorithm enters \stepfive.
Note that this step mimics the swimming behavior of E. coli bacteria, since the algorithm will continue to move along a direction once it is found to be beneficial and until no further improvement can be made. As a result, we term \stepfour \ as ``swim'' and the number of iterations spent in \stepfour \ as ``swim length''.

In fact, {\bf Steps 2 - 4} can be combined into a single expression given by
\begin{align}
\thetabm[k] = \thetabm[k-1] + \beta_{k} \cdot\deltabm[k] \label{eqn:aggregatesteps}
\end{align}
where
\begin{align}
\beta_{k} = \underset{N = 0,\pm 1,\cdots,\pm L.Swim}{\arg \max} \mag(\thetabm[k-1] + N \deltabm[k]) \label{eqn:betak}
\end{align}

Notice that we impose a maximum swim length ``$L.Swim$''.
This maximum swim length is set to prevent situations where the algorithm ``swims'' too far along a direction that only provides slight improvement on the performance. In this case, the expected improvement on the received signal magnitude through swimming is usually less than that achieved by generating new perturbations. It has been verified via simulations that imposing such limit improves overall performances.

\emphb{Step 5 - Stepsize Adjustment}:
For every $L.Held$ iterations, the algorithm adjusts its sampling stepsize $\Delta_{k}$ according to the average swim length during the previous $L.Held$ iterations.
Also, the adjustment made to $\Delta_{k}$ is no less than a factor $\alpha$. In cases where the average swim length is close to zero, this $\alpha$ factor prevents the sampling stepsize $\Delta_{k}$ from becoming too small\footnote{Setting sampling stepsize with values that are too small can sometimes cause random search algorithms to virtually ``trap'' in the local region, and severely degrade algorithms' performance.}.
The stepsize adjustment can be described by
\begin{align}
\Delta_{k} = \Delta_{k-1} \cdot \max\left ( \alpha , \sum_{i = k - L.Held + 1}^{k} \left | \beta_{k} \right | \ / \ {L.Helds} \right )
\end{align}

The overall effect can be seen as the algorithm adaptively increase the sampling stepsize if there were many ``swims'', which indicates that the perturbations generated were small on average, and decrease the sampling stepsize by a factor as small as $\alpha$ if there were only few or no ``swims'', which indicates that the search area is too large.

For iterations that involve no adjustment, the value of $\Delta_{k}$ remains unchanged as in the last iteration.
No matter there is an adjustment or not, the algorithm returns to \steptwo, which marks the end of the current iteration.

\emphb{Stopping criterion}: During each step of BioRARSA, if the received signal magnitude is greater than a target value\footnote{The target value can be based on statistics of the channel, which is assume to be available and known by receiver.}, the algorithm exits and the phase alignment for transmit beamforming is said to be completed.

\subsection{Important Characteristic of BioRARSA}
As will become clear in section~\ref{sec:sim}, 
the convergence behavior of BioRARSA is either insensitive to the choice of parameters at the initialization or to the number of distributed transmitters in the system.
Therefore, all necessary parameters can be pre-coded and  transmission overheads in \stepone~can be eliminated.

Notice that although the algorithm is more complicated than the one-bit adaptive scheme, there is neither extra requirement on the feedback information nor the need for information exchange between distributed transmitters.
The added cost is the computational complexity to execute the algorithm in the microprocessor equipped in each transmitter.
However, energy consumed for such computation is negligible compared with that consumed by transmission.
BioRARSA is hence a solid improvement over existing schemes.

\section{Convergence Analysis}\label{sec:converge}
The analysis of convergence can be greatly simplified via the use of the random search framework proposed in \cite{Chen2010}, which provides a general proof if the reformulated random search algorithm satisfies the following two conditions:
\begin{enumerate}
\item The origin is an interior point of the support of the probability measure $\mu_{n}$ for all n.
\item The objective function $f : \Theta \rightarrow \mathbb{R}$ is continuous and all of its local maxima are global maxima.
\end{enumerate}
These properties jointly provide the sufficient conditions for a local random search algorithm to converge. Therefore, in the ensuing discussions, we reformulate BioRARSA as a random search algorithm and examine these sufficient conditions. Once these conditions are satisfied, we can then apply the framework to show the convergence and linear scalability of BioRARSA.

\subsection{Reformulation and Sufficient Conditions}
Based on \eqref{eqn:aggregatesteps} and \eqref{eqn:betak}, the reformulation is as follows:
\begin{align}
f(\cdot) & = \mag(\cdot)\label{eqn:freform}\\
\mu_{n} &= \text{uni}( \left [ -\Delta_{n} , \Delta_{n} \right ]^{N_{s}} )\label{eqn:munreform}\\
\text{Range} \left \{D(\thetabm,\deltabm[n])\right \} & = \left\{ \thetabm + \Omega \right \} \ \cap \ \Utheta \label{eq:dreform}
\end{align}
where
\begin{align*}
\Utheta &:= \left \{ \omegabm \in \Theta \  | \  \mag(\omegabm) \geq \mag(\thetabm)\right \}\\
\Omega &= \{\hat\omegabm | \hat\omegabm = L.Swim\cdot \omegabm , \omegabm \in \supp(\mu_{n})\}
\end{align*}
Unlike the original one-bit adaptive scheme, the support of $\deltabm[n]$ is extended to $\Omega$ since the ``swim'' step allows the perturbation to reach further by a factor of $L.Swim$.
Although the probability distribution in $\Omega$ is no longer uniform, it is clear that the origin still lies in the interior of $\Omega$. Therefore, the first sufficient condition is satisfied.

The property of $\mag(\cdot)$ has been identified by Lin et al \cite{Lin2006} that all its local maxima are global maxima. This property is important since it ensures that a local random search algorithm can never be trapped in local maxima.
Hence, the second sufficient condition is also satisfied.

\subsection{Proof of Convergence}
As both sufficient conditions are satisfied, we proceed to the proof of convergence with the help of the framework in \cite{Chen2010}. Since the framework guarantees convergence  of the target algorithm once both sufficient conditions are satisfied, the proofs parallel those derived in \cite{Chen2010} and are hence omitted here due to space constraint. The sufficient conditions lead to the following proposition.

\begin{prop}\label{prop:pos_improve}
For any given $\thetabm \in \Theta \setminus R_{\epsilon}$ and $\Delta > 0$, there exists $\gamma >0$ and $0< \eta \leq 1$ such that
\begin{equation}
\nonumber
\textrm{Pr} \left[ f(\thetabm+\deltabm) - f(\thetabm) \geq \gamma \right] \geq \eta
\end{equation}
where $\deltabm$ is a random vector generated with the probability measure $\mu_n$, $R_{\epsilon}$ is the $\epsilon$-convergence region as stated in \cite{Lin2006}, which is defined as $R_{\epsilon} := \left\{ \thetabm \in \Theta : f(\thetabm) > f\left(\thetabm^{\star}\right) -\epsilon \right\}$, and $\thetabm^{\star}$ is a global maximum point.
\end{prop}

\emph{Proposition~\ref{prop:pos_improve}} states that there is always a non-zero probability to improve the received signal magnitude within each iteration before the stopping criterion is satisfied. Based on this, the random search framework provides the following theorem
\begin{theorem}\label{thm:converge}
The sequence $\left\{\thetabm[n]\right\}_{n=1}^\infty$ converges in probability, i.e., given $\epsilon>0$,
\begin{equation}\nonumber
\lim_{n \rightarrow \infty} \textrm{Pr}\left[ \thetabm[n] \in R_{\epsilon} \right] = 1
\end{equation}
\end{theorem}

As the sequence converges to $R_{\epsilon}$, the received signal magnitude becomes greater or equal to $\mag(\thetabm^{\star}) - \epsilon$. By setting $\epsilon$ infinitely small, we prove the convergence of BioRARSA.

\subsection{Linear Scalability}
BioRARSA inherits the same desirable feature of linear scalability from its predecessor. This can also be proven by using the random search framework.
\begin{theorem}
The average time of convergence scales linearly with $N_{s}$. That is, there exists $k < \infty$, such that
\[
\mathrm{E}\left[ \mag(\thetabm[n]) \right] \geq \mag(\thetabm^{\star}) - \epsilon, \quad \forall n \geq k\cdot N_{s}.
\]
\end{theorem}

The detailed proof is omitted due to space constraint, but a brief description is given here:
From the proof in \cite{Lin2006}, it has been shown that the reformulated random search algorithm scales linearly with $N_{s}$. Also, notice that the total number of transmissions of BioRARSA is upper bounded by $L.Swim$ times the total number of transmissions of the reformulated algorithm described by \eqref{eqn:freform}-\eqref{eq:dreform}.
Since the reformulated algorithm is linearly scalable by the random search framework \cite{Lin2006}, BioRARSA must also be linearly scalable.

\section{Simulation result}\label{sec:sim}
\begin{figure}[t]
\psfrag{inidelta3}[lc][lc][0.7]{$\Delta_{0}$ = 3}
\psfrag{inidelta9}[lc][lc][0.7]{$\Delta_{0}$ = 9}
\psfrag{newalgorithm}[lc][lc][0.6]{BioRARSA}
\psfrag{reversetracking____}[lc][lc][0.6]{Reverse Tracking}
\psfrag{onebitfeedback}[lc][lc][0.6]{One-bit Adaptive}
\psfrag{reversetrackwswim}[lc][lc][0.6]{RT with Swim}
\psfrag{targetmagnitude}[lc][lc][0.6]{Target Magnitude}
\psfrag{title}{}
\centering{
\psfrag{xlab}[cc][cc][0.7]{Transmission Iterations}
\psfrag{ylab}[cc][cc][0.7]{Received Magnitude}
\includegraphics[width=3.4in]{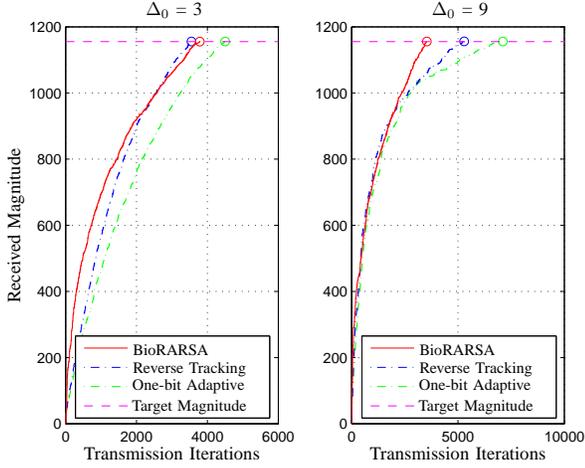}
\caption{Trajectories of received signal magnitude of a 800-node system.}
\label{fig:magtrail}
}
\end{figure}
\begin{figure}[t]
\psfrag{newalgorithm}[lc][lc][0.6]{BioRARSA}
\psfrag{reversetracking______}[lc][lc][0.6]{Reverse Tracking}
\psfrag{onebitfeedback}[lc][lc][0.6]{One-bit Adaptive}
\psfrag{reversetrackwswim}[lc][lc][0.6]{RT with Swim}
\psfrag{targetmagnitude}[lc][lc][0.6]{Target Magnitude}
\psfrag{title}{}
\psfrag{xlab}[cc][cc][0.7]{Initial Sampling Stepsize, $\Delta_{0}$}
\psfrag{ylab}[cc][cc][0.7]{Average Number of Transmissions}
\centering{
\includegraphics[width=3.4in]{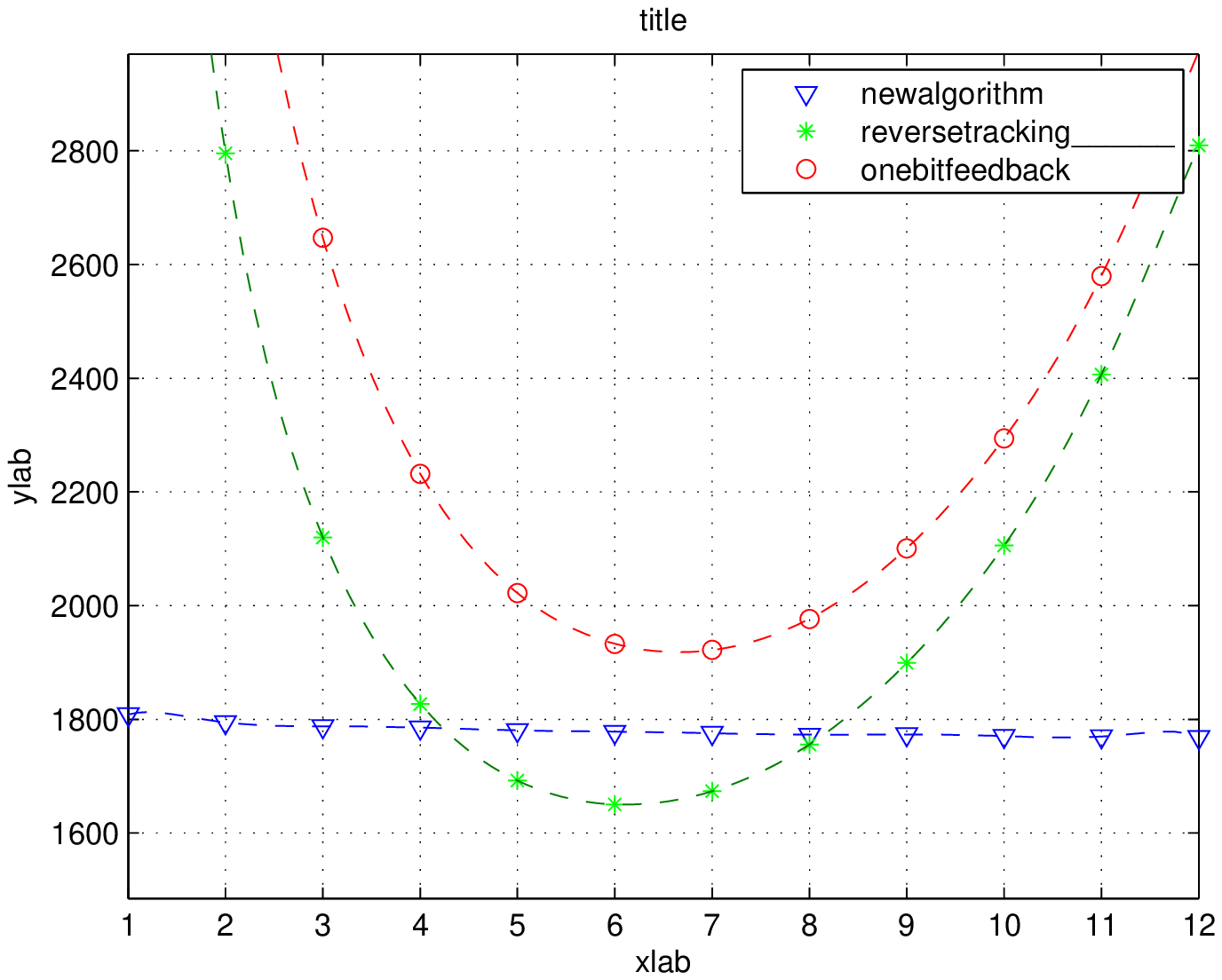}
\caption{Performance comparison of 400-node systems.}
\label{fig:perfcomp200}
}
\end{figure}

\begin{figure}[t]
\psfrag{newalgorithm}[lc][lc][0.6]{BioRARSA}
\psfrag{reversetracking______}[lc][lc][0.6]{Reverse Tracking}
\psfrag{onebitfeedback}[lc][lc][0.6]{One-bit Adaptive}
\psfrag{reversetrackwswim}[lc][lc][0.6]{RT with Swim}
\psfrag{targetmagnitude}[lc][lc][0.6]{Target Magnitude}
\psfrag{title}{}
\psfrag{xlab}[cc][cc][0.7]{Initial Sampling Stepsize, $\Delta_{0}$}
\psfrag{ylab}[cc][cc][0.7]{Average Number of Transmissions}
\centering{
\includegraphics[width=3.4in]{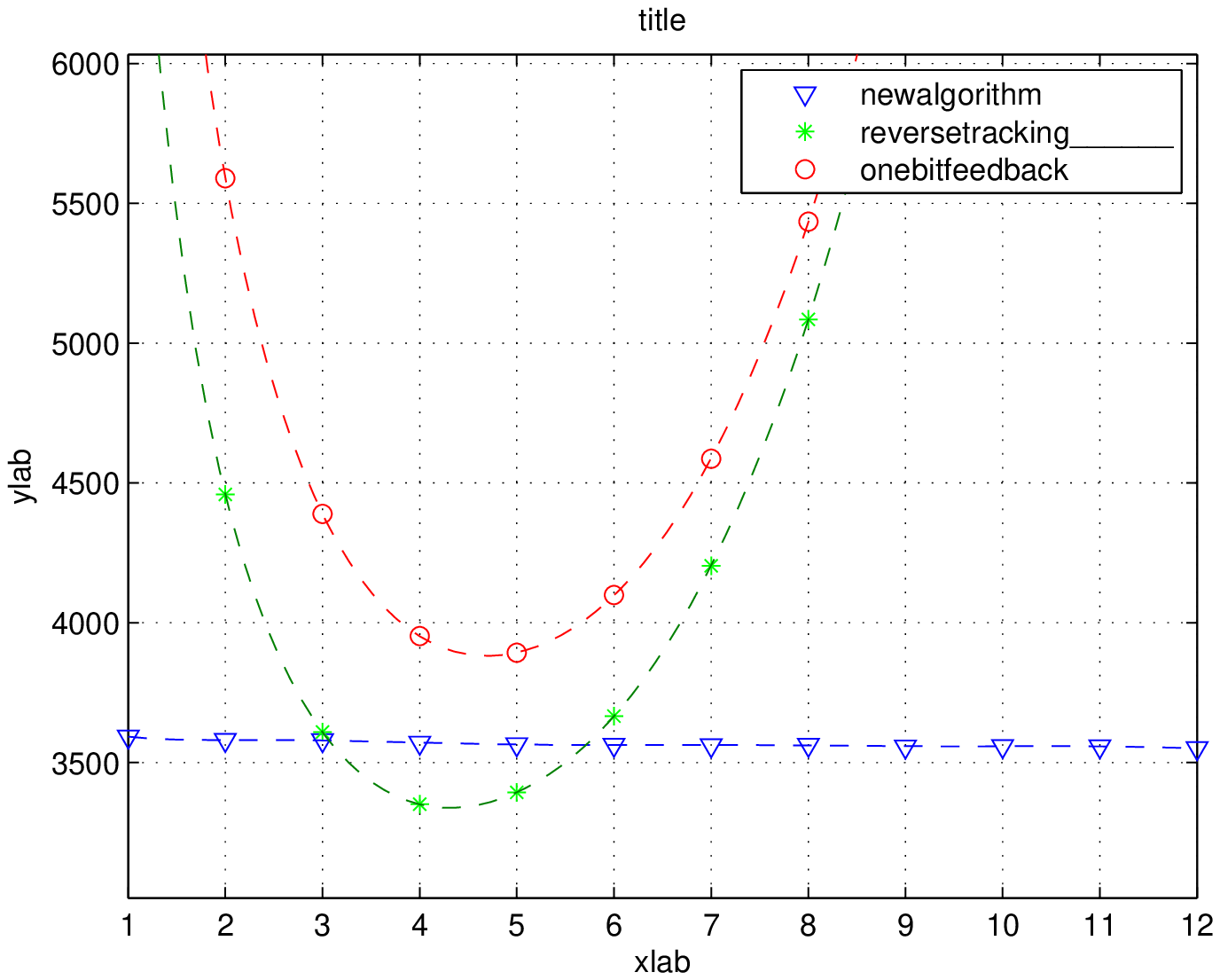}
\caption{Performance comparison of 800-node systems.}
\label{fig:perfcomp800}
}
\vfill
\end{figure}

\begin{figure}[t]
\psfrag{targ__mag06}[lc][lc][0.6]{$C_{stop} = 0.6$}
\psfrag{targ__mag07}[lc][lc][0.6]{$C_{stop} = 0.7$}
\psfrag{targ__mag08}[lc][lc][0.6]{$C_{stop} = 0.8$}
\psfrag{targ__mag09}[lc][lc][0.6]{$C_{stop} = 0.9$}
\psfrag{xlab}[cc][cc][0.7]{Number of Transmitters in the system}
\psfrag{ylab}[cc][cc][0.7]{Average Number of Transmissions}
\includegraphics[width=3.4in]{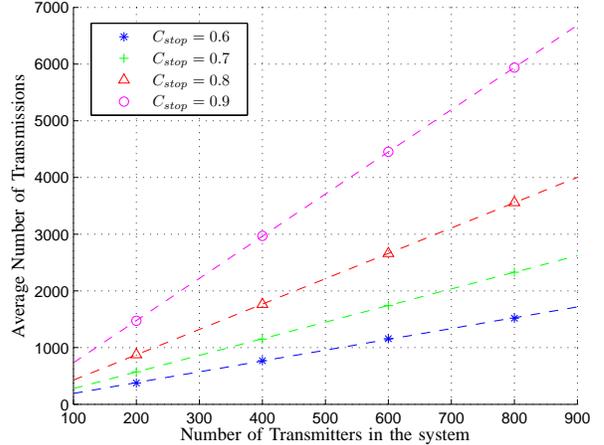}
\caption{Linear scalability of BioRARSA.}
\label{fig:linscale}
\end{figure}

\begin{figure}[t]
\psfrag{trailofdelta____}[lc][lc][0.6]{$\Delta_n$ Evolution}
\psfrag{init}[lc][lc][0.6]{Initial $\Delta_n$}
\psfrag{end}[lc][lc][0.6]{Final $\Delta_n$}
\psfrag{xlab}[cc][cc][0.7]{Stepsize adjustment iterations}
\psfrag{ylab}[cc][cc][0.7]{Degrees}
\psfrag{inidelta3}[cc][cc][0.7]{Initial Stepsize $\Delta_{0}$ = 3}
\psfrag{inidelta9}[cc][cc][0.7]{Initial Stepsize $\Delta_{0}$ = 9}
\centering{
\includegraphics[width=3.4in]{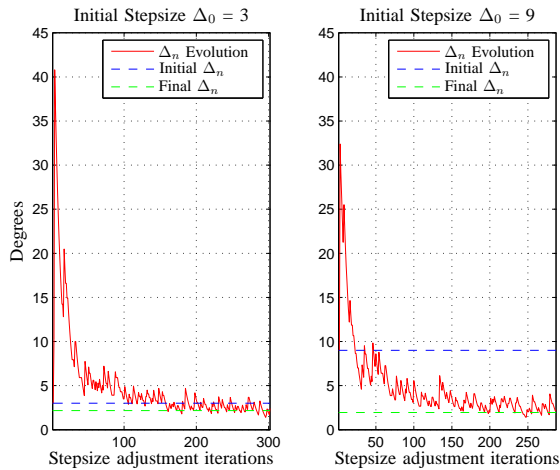}
\caption{Evolutions of sampling stepsizes in a 800-node system.}
\label{fig:deltatrail}
}
\end{figure}

\begin{table}[t]
\centering
\begin{threeparttable}[t]
\renewcommand{\arraystretch}{1.3}
\caption{Performance Gain of BioRARSA with Respect to\newline The Reverse Tracking Scheme ( in \%)}
\label{tab:perfcomp}
\centering
\begin{tabular}[t]{c|rrrrr}
\backslashbox[10mm]{$\Delta_{0}$}{\# Nodes}& 200~~~ & 400~~~ & 600~~~ & 800~~~ & \\
\hline
$1^{\circ}$&   72.3 \%&   63.5 \%&   57.1 \%&   51.9 \%&\\
$3^{\circ}$&   31.5 \%&   15.7 \%&    6.6 \%&    0.8 \%&\\
$5^{\circ}$&    6.2 \%&   -5.2 \%&   -7.4 \%&   -5.1 \%&\\
$7^{\circ}$&   -6.4 \%&   -6.1 \%&    3.4 \%&   15.3 \%&\\
$9^{\circ}$&   -8.7 \%&    6.6 \%&   26.8 \%&   44.9 \%&\\
$11^{\circ}$&   -3.1 \%&   26.5 \%&   52.5 \%&   70.9 \%&\\
\hline
Average Gain   &15.3 \%&   16.8 \%&   23.1 \%&   29.8 \%&\\
\end{tabular}
\end{threeparttable}
\end{table}

Numerical results presented in this section are conducted using Monte Carlo simulations. Each point in Fig.~\ref{fig:perfcomp200} to \ref{fig:linscale} and values in Table~\ref{tab:perfcomp} are averaged over $10^{4}$ random instances that consist of
$10^{2}$  pre-generated channel realization $\{h_{i}\}$ each with $10^{2}$ independent random perturbation sequences $\{ \delta_{i}[n]\}$. Through simulations, we show the performance advantage
and the robustness of BioRARSA, and proofs derived in Sec.~\ref{sec:converge} are also validated.

The performance is measured by the average number of simultaneous transmissions from distributed transmitters to the receiver before reaching the stopping criterion. We set $L.Helds$ and $L.Swim$ to 5 for these values are optimal for most systems. Simulation results that justify these choices of $L.Helds$ and $L.Swim$ are omitted due to space constraint.

To demonstrate the robustness of BioRARSA with respect to the initial choice of sampling stepsize $\Delta_{0}$, we first present two sets of trajectories of received signal magnitude traversed by the randomized algorithms in Fig.~\ref{fig:magtrail}. Each magnitude trajectory corresponds to a realization of the beamforming coefficients recorded during the phase alignment process using the stated schemes, and $\Delta_{0}$ of each set is set to be $3^{\circ}$ and $9^{\circ}$, respectively.
These trajectories confirm that BioRARSA is insensitive to the initial choice of $\Delta_{0}$, while the number of transmissions in the reverse tracking scheme and the one-bit adaptive scheme increase nearly $60\%$ when  $\Delta_{0}$ increase from $3^{\circ}$ to $9^{\circ}$. Note that we have generated these trajectories repeatedly for different realizations and obtained similar behaviors. For all realizations, all three algorithms converge without exception. This also validates our proof in Sec.~\ref{sec:converge}.

Aside from its robust nature, the overall performance of BioRARSA is competitive compared with existing schemes.
From Fig.~\ref{fig:perfcomp200} and~\ref{fig:perfcomp800}, it can be identified that the one-bit adaptive scheme is completely outperformed by the reverse tracking scheme and BioRARSA, and there exists a small set of $\Delta_{0}$ where reverse tracking outperforms BioRARSA if the reverse tracking scheme happens to choose the stepsizes that are close to the optimal values.
However, the optimal choice of $\Delta_{0}$ is closely related to the number of distributed transmitters in the system, which is often difficult to track in a distributed network. On the contrary, BioRARSA is capable of providing near optimal performance without any prior knowledge of the system, making it more suitable for distributed networks.

To provide a more detailed performance comparison between the reverse tracking scheme and BioRARSA, the normalized performance gain $
\left ( N_{RT} - N_{BioRARSA} \right ) / N_{RT} \cdot 100 \% $ is presented in Table~\ref{tab:perfcomp}, where $N_{RT}$ and ${N_{BioRARSA}}$ are the average number of transmissions with respect to each scheme. In this setting, positive gain indicates that BioRARSA outperforms the reverse tracking scheme, and vice versa.
By comparing the overall performance, we can see that the insensitivity to the initial choice of parameters allows BioRARSA to achieve better performance on a wide range of initial sampling stepsize compared with the reverse tracking scheme.
In certain cases, the performance gain of BioRARSA over reverse tracking is as large as $29.8\%$ on average.

In section~\ref{sec:converge} it was proven that BioRARSA scales linearly with respect to the system size, and this characteristic is further validated with the simulation  shown in Fig.~\ref{fig:linscale}.
Each line in the figure corresponds to different stopping criteria where the algorithm is stopped when $\mag(\thetabm[n]) \geq C_{stop} \cdot \mag(\thetabm^{\star})$, and $\mag(\thetabm^{\star})$ is the maximum achievable received signal magnitude as defined in Sec.~\ref{sec:converge}.
From this, it is clear that BioRARSA is indeed linearly scalable with target magnitude ranging from 60\% to 90\% of $\mag(\thetabm^{\star})$.

In Fig.~\ref{fig:deltatrail}, the evolutions of sampling stepsize $\Delta_n$ are presented, where the simulation setup is the same as those in Fig.~\ref{fig:magtrail}.
We observe that the adjusted sampling stepsize first increases drastically and then decreases gradually as the beamforming coefficients approach their optimal values, which confirms with the concept of ``optimal relative stepsize\footnote{\emph{``Relative stepsize''} is defined as the ratio of the current sampling stepsize to the distance of the current point from the optimal point. In \cite{Schrack1976}, it has been shown that by using optimal relative stepsize in every iteration of a random search algorithm, optimal performance can be achieved.}'' as proposed in \cite{Schrack1976}. Although only two samples were presented, this phenomenon was observed on most recorded samples.

\section{conclusion}\label{sec:con}
In this work, we propose a bio-inspired distributed beamforming scheme, BioRARSA, based on our understanding of an existing one-bit adaptive scheme via a random search framework, and further inspired by a heuristic algorithm that mimics the bacterial foraging behavior.

Our numerical simulations have validated the proofs derived in Sec.~\ref{sec:converge}, and shown that BioRARSA is more robust and efficient than the one-bit adaptive scheme uniformly and the reverse tracking scheme on average.
This improved performance is achieved via the ``swim'' behavior that adjusts the sampling stepsize adaptively, and hence allowing BioRARSA to be robust against system uncertainties.

In the future, we are interested in further examination of the robustness of BioRARSA under more practical systems.
We would also like to improve our framework further such that it can also provide analysis of the rate of convergence.
\vfill
\bibliographystyle{IEEEtran}
\bibliography{ref}

\begin{thebibliography}{1}
\providecommand{\url}[1]{#1}
\csname url@samestyle\endcsname
\providecommand{\newblock}{\relax}
\providecommand{\bibinfo}[2]{#2}
\providecommand{\BIBentrySTDinterwordspacing}{\spaceskip=0pt\relax}
\providecommand{\BIBentryALTinterwordstretchfactor}{4}
\providecommand{\BIBentryALTinterwordspacing}{\spaceskip=\fontdimen2\font plus
\BIBentryALTinterwordstretchfactor\fontdimen3\font minus
  \fontdimen4\font\relax}
\providecommand{\BIBforeignlanguage}[2]{{%
\expandafter\ifx\csname l@#1\endcsname\relax
\typeout{** WARNING: IEEEtran.bst: No hyphenation pattern has been}%
\typeout{** loaded for the language `#1'. Using the pattern for}%
\typeout{** the default language instead.}%
\else
\language=\csname l@#1\endcsname
\fi
#2}}
\providecommand{\BIBdecl}{\relax}
\BIBdecl

\bibitem{Mudumbai2006}
R.~Mudumbai, B.~Wild, U.~Madhow, and K.~Ramch, ``Distributed beamforming using
  1 bit feedback: From concept to realization,'' in \emph{Allerton Conference
  on Communication, Control, and Computing}, 2006.

\bibitem{Lin2006}
C.~Lin, V.~V. Veeravalli, and S.~P. Meyn, ``A random search framework for
  convergence analysis of distributed beamforming with feedback,'' \emph{IEEE
  Trans. Information Theory}, vol.~56, no.~12, Dec 2010.

\bibitem{Bucklew2008}
J.~Bucklew and W.~Sethares, ``Convergence of a class of decentralized
  beamforming algorithms,'' \emph{IEEE Trans. Signal Processing}, vol.~56,
  2008.

\bibitem{Chen2010}
C.~C. Chen, C.~S. Tseng, and C.~Lin, ``A general proof of convergence for
  adaptive distributed beamforming schemes,'' in \emph{International Conference
  on Acoustics, Speech and Signal Processing}, 2011.

\bibitem{Schrack1976}
G.~Schrack and M.~Choit, ``Optimized relative step size random searches,''
  \emph{Mathematical Programming}, vol.~10, 1976.

\bibitem{Solis1981}
F.~J. Solis and R.~J.-B. Wets, ``Minimization by random search techniques,''
  \emph{Mathematics of Operations Research}, vol.~6, 1981.

\bibitem{Passino2002}
K.~Passino, ``Biomimicry of bacterial foraging for distributed optimization
  andcontrol,'' \emph{IEEE Control Systems Magazine}, vol.~22, 2002.

\end{thebibliography}
\end{document}